\documentclass[%
reprint,
 amsmath,amssymb,
aps,
pra,
]{revtex4-1}

\usepackage{graphicx}
\usepackage{dcolumn}
\usepackage{bm}

\begin{document}

\preprint{APS/123-QED}

\title{MAICRM: A general model for rapid simulation of hot dense plasma spectra}%

\author{Xiaoying Han}%
\email{han\_xiaoying@iapcma.ac.cn}
\author{Lingxiao Li}%
\author{Zhensheng Dai}%
\author{Wudi Zheng}%
\affiliation{Institute of Applied Physics and Computational Mathematics, Beijing 100094, China}%

\date{\today}

\begin{abstract}
This work is to continue the development of the general model, Multi-Average Ion Collisional-Radiative Model (MAICRM), to calculate the plasma spectral properties of hot dense plasmas. In this model, an average ion is used to characterize the average orbital occupations and the total populations of the configurations within a single charge state. The orbital occupations and population of the average ion are obtained by solving two sets of rate equations sequentially and iteratively. The calculated spectra of Xe and Au plasmas under different plasma conditions are in good agreement with the DCA/SCA calculations while the computational cost is much lower.
\end{abstract}

\pacs{Valid PACS appear here}
\maketitle

\section{Introduction}
To accurately model the properties of matter under non-local thermodynamic equilibrium (NLTE) conditions is critical to simulate the evolution of plasmas produced in inertial confinement fusion (ICF), magnetic fusion and other laboratory experiments as well as to diagnose the states of these systems by simulating the spectra. For most NLTE simulations the collisional-radiative (CR) model\cite{Bates1962} is used. Book\cite{Ralchenkobook} reviewed the history and the recent developments of modern methods employed in CR modeling. In general, NLTE atomic models can be classified into two kinds: one is to characterize the plasma properties by experientially choosing the important atomic levels, such as the detailed configuration accounting (DCA), superconfiguration accounting (SCA) or hybrid models\cite{SOSA1991,STA1989}; the other one is average-atom (AA) model which characterizes the plasma simply by the orbital occupations of one ion, which is the average of all the atomic levels. Although much effort has gone into developing the NLTE simulations, such as extended or reduced DCA models\cite{RDCA2008,RDCA2009,Scott2010}, the questions of which (or how many) states and transitions to include in an atomic model and what degree of averaging to use as well as the computational cost are still critical and unsolved completely for an on-line detailed atomic calculation incorporated in radiation-hydrodynamic simulations.

In our work\cite{MAICRM-1} we reported a general atomic model MAICRM (Multi-Average Ion Collisional-Radiative Model) to rapidly simulate the average ionization $\langle Z\rangle$ and the charge state distribution (CSD) of hot dense plasmas (labeled as paper I in the following). The main thought of MAICRM is using an average ion to characterize the plasma properties of all the configurations at one charge state. The orbital occupation of an average ion is the average of the orbital occupations of all the configurations at one charge state and its population is the sum of the configuration populations. The idea of using multi-average ions to simulate plasma has been presented before\cite{Kiyokawa2014,Itoh1987} but the procedures to get the orbital occupations and the populations of the average ions are different. Ref.\cite{Kiyokawa2014} proposed a multi-average ion model for LTE plasma, in which a Fermi distribution for the orbital occupation numbers of each average ion is assumed and the populations of the ions are calculated self-consistently through minimizing the free energy of the whole system established by the finite temperature density functional theory. Ref.\cite{Itoh1987} reported a hybrid-atom model, which estimated the excited level populations approximately by quasisteady equations of bound-bound transitions in an ion of charge $Z$. The normalized populations of excited levels constructed from the most populated ion are scaled and used for other ions at different charge states. Then the rate coefficients for each average ion are constructed by the summation over all excited states weighted by the level populations and the populations of the ions are obtained by solving the rate equations only including ionization/recombination processes.

Differing from the former work, our model calculates the orbital occupation numbers by solving the rate equations involving bound-bound (excitation/de-excitation) transitions, and the populations by solving the rate equations involving ionization/recombination processes with fixed occupation numbers. These rate equations derive from the rate equations of detailed configurations. During the transformation from the detailed configuration rate equations into two sets of rate equations for average ions, two assumptions are adopted: one is the single orbital rate coefficients (without the occupation and hole numbers) are only the functions of the charge state; the other one is the coupled rate equations of occupation numbers and populations of average ions are separated.

Paper I has shown the calculated $\langle Z\rangle$ and CSD of Fe, Xe and Au plasmas under various plasma conditions by MAICRM codes agree with the experimental and DCA/SCA results. This work will continue the development of the method to calculate the plasma spectral properties of hot dense plasmas, including the formulas to calculate the position and width of transition array for average ions. Using the developed codes, our calculated single orbital transition spectra agree with those of super transition arrays (STA) model with a single supershell for each charge state\cite{STA1989}. The present calculated emissivity spectra of Xe and Au plasmas agree well with the detailed SCRAM\cite{Scott2010} and DCA results\cite{RDCA2009}. Since for each ionic stage only one configuration is calculated in MAICRM, the computational cost is lower than SCA/DCA by magnitudes due to their hundreds or thousands superconfigurations/configurations at each charge state.

\section{Theoretical method}

\subsection{Rate equations}
Since the detailed description of the derivation of the rate equations in MAICRM has been presented in paper I, here only a brief outline will be given. The essential of MAICRM is using an average ion to describe the properties of all the configurations at one charge state. More specifically, for an average ion $\mathbf{\Lambda_{n_{e}}}$, its occupation number $\mathbf{\Omega}^{\mathbf{\Lambda_{n_{e}}}}_{n}$ is the average of the configuration's orbital occupation numbers $\{w^{ \mathbf{K_{n_{e}}}}_{n}\}$ of one charge and its population $P_{\mathbf{\Lambda_{n_{e}}}}$ is the summation of the configuration populations $\{P_{\mathbf{K_{n_{e}}}}\}$, namely
\begin{eqnarray}
\label{eq:MAIAPPRO}
&&\mathbf{\Omega}^{\mathbf{\Lambda_{n_{e}}}}_{n} = \frac{\sum_{ \mathbf{K_{n_{e}}}}P_{\mathbf{K_{n_{e}}}}w^{ \mathbf{K_{n_{e}}}}_{n}}{\sum_{ \mathbf{K_{n_{e}}}}P_{\mathbf{K_{n_{e}}}}} \nonumber \\
&&P_{\mathbf{\Lambda_{n_{e}}}}  = \sum_{ \mathbf{K_{n_{e}}}}P_{\mathbf{K_{n_{e}}}}
\end{eqnarray}
Here $\mathbf{K_{n_{e}}}$ labels a configuration of $\mathbf{n_{e}}$ bound electrons.

By summing the rate equations of orbital occupations of the configurations multiplied by the configuration population, we get the rate equations of $P_{\mathbf{\Lambda_{n_{e}}}}$ and $\mathbf{\Omega}^{\mathbf{\Lambda_{n_{e}}}}_{n}$
\begin{widetext}
\begin{eqnarray}
\label{eq:MAIRE2}
\frac{ \mathbf{d} (P_{\mathbf{\Lambda_{n_{e}}}}\mathbf{\Omega}^{\mathbf{\Lambda_{n_{e}}}}_{n})}{\mathbf{d}t}&= &P_{\mathbf{\Lambda_{n_{e}}}} \frac{  \mathbf{d}\mathbf{\Omega}^{\mathbf{\Lambda_{n_{e}}}}_{n} }{\mathbf{d}t}+\mathbf{\Omega}^{\mathbf{\Lambda_{n_{e}}}}_{n}\frac{  \mathbf{d}P_{\mathbf{\Lambda_{n_{e}}}} }{\mathbf{d}t}  \nonumber  \\
&=& P_{\mathbf{\Lambda_{n_{e}}}}\sum_{l}(\mathbf{g}_{n} - \mathbf{\Omega}^{\mathbf{\Lambda_{n_{e}}}}_{n}) \mathbf{\Omega}^{\mathbf{\Lambda_{n_{e}}}}_{l} \mathbf{R}^{ \mathbf{E/D} }_{ \mathbf{n_{e}},l\rightarrow n} - P_{\mathbf{\Lambda_{n_{e}}}}\sum_{l}(\mathbf{g}_{l} - \mathbf{\Omega}^{\mathbf{\Lambda_{n_{e}}}}_{l}) \mathbf{\Omega}^{\mathbf{\Lambda_{n_{e}}}}_{n} \mathbf{R}^{ \mathbf{E/D} }_{ \mathbf{n_{e}},n\rightarrow l} \nonumber  \\
&&+ \mathbf{\Omega}^{\mathbf{\Lambda_{n_{e}}}}_{n} P_{\mathbf{\Lambda_{n_{e}+1}}} \sum_{i}\mathbf{\Omega}^{\mathbf{\Lambda_{n_{e}+1}}}_{i}\mathbf{R}^{ \mathbf{I} }_{ \mathbf{n_{e}+1},i\uparrow }
  + \mathbf{\Omega}^{\mathbf{\Lambda_{n_{e}}}}_{n} P_{\mathbf{\Lambda_{n_{e}-1}}} \sum_{i}(\mathbf{g}_{i} - \mathbf{\Omega}^{\mathbf{\Lambda_{n_{e}-1}}}_{i})\mathbf{R}^{ \mathbf{R} }_{ \mathbf{n_{e}-1},i\downarrow }  \nonumber  \\
&&- \mathbf{\Omega}^{\mathbf{\Lambda_{n_{e}}}}_{n} P_{\mathbf{\Lambda_{n_{e}}  }} \sum_{i}\mathbf{\Omega}^{\mathbf{\Lambda_{n_{e}  }}}_{i}\mathbf{R}^{ \mathbf{I} }_{ \mathbf{n_{e}  },i\uparrow }
  - \mathbf{\Omega}^{\mathbf{\Lambda_{n_{e}}}}_{n} P_{\mathbf{\Lambda_{n_{e}}  }} \sum_{i}(\mathbf{g}_{i} - \mathbf{\Omega}^{\mathbf{\Lambda_{n_{e}  }}}_{i})\mathbf{R}^{ \mathbf{R} }_{ \mathbf{n_{e}  },i\downarrow }  \nonumber  \\
&&+ \mathbf{\Omega}^{\mathbf{\Lambda_{n_{e}}}}_{n} P_{\mathbf{\Lambda_{n_{e}+1}}}\sum_{l,u,i}(\mathbf{g}_{l}-\mathbf{\Omega}^{\mathbf{\Lambda_{n_{e}+1}}}_{l})(\mathbf{\Omega}^{\mathbf{\Lambda_{n_{e}+1}}}_{u}-\delta_{ui})\mathbf{\Omega}^{\mathbf{\Lambda_{n_{e}+1}}}_{i}
\mathbf{R}^{ \mathbf{AI} }_{ \mathbf{n_{e}+1},u \rightarrow l, i\uparrow}  \nonumber  \\
&& + \mathbf{\Omega}^{\mathbf{\Lambda_{n_{e}}}}_{n} P_{\mathbf{\Lambda_{n_{e}-1}}}\sum_{l,u,i}\mathbf{\Omega}^{\mathbf{\Lambda_{n_{e}-1}}}_{l}(\mathbf{g}_{u}-\mathbf{\Omega}^{\mathbf{\Lambda_{n_{e}-1}}}_{u}-\delta_{ui})(\mathbf{g}_{i}-\mathbf{\Omega}^{\mathbf{\Lambda_{n_{e}-1}}}_{i})
\mathbf{R}^{ \mathbf{EC} }_{ \mathbf{n_{e}-1},l \rightarrow u, i\downarrow}    \nonumber  \\
&& - \mathbf{\Omega}^{\mathbf{\Lambda_{n_{e}}}}_{n} P_{\mathbf{\Lambda_{n_{e}}}}\sum_{l,u,i}(\mathbf{g}_{l}-\mathbf{\Omega}^{\mathbf{\Lambda_{n_{e}}}}_{l})(\mathbf{\Omega}^{\mathbf{\Lambda_{n_{e}}}}_{u}-\delta_{ui})\mathbf{\Omega}^{\mathbf{\Lambda_{n_{e}}}}_{i}
\mathbf{R}^{ \mathbf{AI} }_{ \mathbf{n_{e}},u\rightarrow l, i\uparrow}  \nonumber  \\
&& - \mathbf{\Omega}^{\mathbf{\Lambda_{n_{e}}}}_{n} P_{\mathbf{\Lambda_{n_{e}}}}\sum_{l,u,i}\mathbf{\Omega}^{\mathbf{\Lambda_{n_{e}}}}_{l}(\mathbf{g}_{u}-\mathbf{\Omega}^{\mathbf{\Lambda_{n_{e}}}}_{u}-\delta_{ui})(\mathbf{g}_{i}-\mathbf{\Omega}^{\mathbf{\Lambda_{n_{e}}}}_{i})
\mathbf{R}^{ \mathbf{EC} }_{ \mathbf{n_{e}},l\rightarrow u, i\downarrow}
\end{eqnarray}
\end{widetext}
Here $\mathbf{R}^{\mathbf{E/D}}$ are the excitation/de-excitation rate coefficients including the electron collision excitation/de-excitation and the photon excitation and spontaneous \& stimulated emissions processes. $\mathbf{R}^{\mathbf{I/R}}$ are the ionization/recombination rate coefficients including the electron collision ionization/three-body recombination and photon ionization/radiative recombination processes. $\mathbf{R}^{\mathbf{AI/EC}}$ are the autoionization and electron capture rate equations. All the single-orbital rate coefficients are only the functions of the charge stage, whose calculation formulas are listed in appendix A of paper I.
Here is our first assumption that besides the single-orbital rate coefficients the single-orbital transition energies and oscillator strengths (without the occupation numbers) are the same for the configurations of the same charge.

Considering the bound-bound transition processes are usually faster than the bound-free atomic processes, we decouple Eq.(\ref{eq:MAIRE2}) into two rate equations:
\begin{widetext}
\begin{eqnarray}
\label{eq:MAIRE3}
\frac{ \mathbf{d}\mathbf{\Omega}^{\mathbf{\Lambda_{n_{e}}}}_{n} }{\mathbf{d}t} &=&
      (\mathbf{g}_{n} - \mathbf{\Omega}^{\mathbf{\Lambda_{n_{e}}}}_{n}) \sum_{l}\mathbf{\Omega}^{\mathbf{\Lambda_{n_{e}}}}_{l} \mathbf{R}^{ \mathbf{E/D} }_{ \mathbf{n_{e}},l\rightarrow n} - \mathbf{\Omega}^{\mathbf{\Lambda_{n_{e}}}}_{n}                    \sum_{l}(\mathbf{g}_{l}-\mathbf{\Omega}^{\mathbf{\Lambda_{n_{e}}}}_{l}) \mathbf{R}^{ \mathbf{E/D} }_{ \mathbf{n_{e}},n\rightarrow l} \\
\label{eq:MAIRE4}
\frac{  \mathbf{d}P_{\mathbf{\Lambda_{n_{e}}}} }{\mathbf{d}t} &=&
    P_{\mathbf{\Lambda_{n_{e}+1}}}\sum_{i}\mathbf{\Omega}^{\mathbf{\Lambda_{n_{e}+1}}}_{i}\mathbf{R}^{ \mathbf{I} }_{ \mathbf{n_{e}+1},i\uparrow }
  + P_{\mathbf{\Lambda_{n_{e}-1}}}\sum_{i}(\mathbf{g}_{i} - \mathbf{\Omega}^{\mathbf{\Lambda_{n_{e}-1}}}_{i})\mathbf{R}^{ \mathbf{R} }_{ \mathbf{n_{e}-1},i\downarrow }  \nonumber  \\
&&- P_{\mathbf{\Lambda_{n_{e}  }}}\sum_{i}\mathbf{\Omega}^{\mathbf{\Lambda_{n_{e}  }}}_{i}\mathbf{R}^{ \mathbf{I} }_{ \mathbf{n_{e}  },i\uparrow }
  - P_{\mathbf{\Lambda_{n_{e}  }}}\sum_{i}(\mathbf{g}_{i} - \mathbf{\Omega}^{\mathbf{\Lambda_{n_{e}  }}}_{i})\mathbf{R}^{ \mathbf{R} }_{ \mathbf{n_{e}  },i\downarrow }  \nonumber  \\
&&+ P_{\mathbf{\Lambda_{n_{e}+1}}}\sum_{l,u,i}(\mathbf{g}_{l}-\mathbf{\Omega}^{\mathbf{\Lambda_{n_{e}+1}}}_{l})(\mathbf{\Omega}^{\mathbf{\Lambda_{n_{e}+1}}}_{u}-\delta_{ui})\mathbf{\Omega}^{\mathbf{\Lambda_{n_{e}+1}}}_{i}
\mathbf{R}^{ \mathbf{AI} }_{ \mathbf{n_{e}+1},u\rightarrow l, i\uparrow}  \nonumber  \\
&&+P_{\mathbf{\Lambda_{n_{e}-1}}}\sum_{l,u,i}\mathbf{\Omega}^{\mathbf{\Lambda_{n_{e}-1}}}_{l}(\mathbf{g}_{u}-\mathbf{\Omega}^{\mathbf{\Lambda_{n_{e}-1}}}_{u}-\delta_{ui})(\mathbf{g}_{i}-\mathbf{\Omega}^{\mathbf{\Lambda_{n_{e}-1}}}_{i})
\mathbf{R}^{ \mathbf{EC} }_{ \mathbf{n_{e}-1},l\rightarrow u, i\downarrow }    \nonumber  \\
&& -P_{\mathbf{\Lambda_{n_{e}}}}\sum_{l,u,i}(\mathbf{g}_{l}-\mathbf{\Omega}^{\mathbf{\Lambda_{n_{e}}}}_{l})(\mathbf{\Omega}^{\mathbf{\Lambda_{n_{e}}}}_{u}-\delta_{ui})\mathbf{\Omega}^{\mathbf{\Lambda_{n_{e}}}}_{i}
\mathbf{R}^{ \mathbf{AI} }_{ \mathbf{n_{e}},u\rightarrow l, i\uparrow}  \nonumber  \\
&& -P_{\mathbf{\Lambda_{n_{e}}}}\sum_{l,u,i}\mathbf{\Omega}^{\mathbf{\Lambda_{n_{e}}}}_{l}(\mathbf{g}_{u}-\mathbf{\Omega}^{\mathbf{\Lambda_{n_{e}}}}_{u}-\delta_{ui})(\mathbf{g}_{i}-\mathbf{\Omega}^{\mathbf{\Lambda_{n_{e}}}}_{i})
\mathbf{R}^{ \mathbf{EC} }_{ \mathbf{n_{e}},l\rightarrow u, i\downarrow}
\end{eqnarray}
\end{widetext}

Eq.(\ref{eq:MAIRE3}) and Eq.(\ref{eq:MAIRE4}) are solved sequentially and iteratively. Firstly, we solve Eq.(\ref{eq:MAIRE3}) to obtain a set of $\{\mathbf{\Omega}_{n}^{\mathbf{\Lambda_{n_{e}}}}\}$ and then solve Eq.(\ref{eq:MAIRE4}) to obtain $\{P_{\mathbf{\Lambda_{n_{e}}}}\}$ with the fixed $\{\mathbf{\Omega}_{n}^{\mathbf{\Lambda_{n_{e}}}}\}$. Then the electron density $N_{e}$ is updated according to the new ion populations $\{P_{\mathbf{\Lambda_{n_{e}}}}\}$ and all the single-orbital rate coefficients $\mathbf{R}$ are recalculated. In succession, Eq.(\ref{eq:MAIRE3}) and Eq.(\ref{eq:MAIRE4}) are solved sequentially like step one. The iteration continues until $N_{e}$ is invariable and the converged $\{P_{\mathbf{\Lambda_{n_{e}}}}\}$ and $\{\mathbf{\Omega}_{n}^{\mathbf{\Lambda_{n_{e}}}}\}$ are obtained. During the iteration, the influence of excitation/de-excitation processes on $\{P_{\mathbf{\Lambda_{n_{e}}}}\}$ are implemented by $\{\mathbf{\Omega}_{n}^{\mathbf{\Lambda_{n_{e}}}}\}$ in Eq.(\ref{eq:MAIRE4}) and the influence of the ionization/recombination processes on $\{\mathbf{\Omega}_{n}^{\mathbf{\Lambda_{n_{e}}}}\}$ are implied by the variance of $N_{e}$. The applicable conditions for decoupling Eq.(\ref{eq:MAIRE2}) is that the mutual influence of $\{P_{\mathbf{\Lambda_{n_{e}}}}\}$ and $\{\mathbf{\Omega}_{n}^{\mathbf{\Lambda_{n_{e}}}}\}$ is weak. In some extreme plasma conditions, the decoupling is inappropriate and should be dealt with specially. For example, in magnetic fusion, which is a very low density plasma condition, the configurations at each charge stage accumulate on the ground state and the excited orbital occupation numbers are tiny, therefore the influence of the ionization/recombination processes on the excited orbital occupations is notable which will lead to a very slow or even unsuccessful converging process for solving the rate eqations.

\subsection{Spectrum}
The $n$th moment of the distribution for a transition array (A$\rightarrow$B) is
\begin{equation}\label{eq:nmoment}
    \mu_{n} =\frac{\sum_{a\in A,b\in B}(E^{\mathbf{K_{n_{e}}}}_{ab})^{n}\mathbf{w}_{ab}}{\mathbf{W}}
\end{equation}
Here $\mathbf{w}_{ab}$, the weight of the a$\rightarrow$b transition, is equal to the transition rate coefficient $P_{\mathbf{K_{n_{e}}}}w_{a}(g_{b}-w_{b})\mathbf{R}_{ab}$ with $g$ and $w$ are the orbital statistic weight and occupation number respectively. $\mathbf{R}_{ab}$ is the single-orbital rate coefficient. $\mathbf{W}=\sum_{a,b}\mathbf{w}_{ab}$.

The position and width of array A$\rightarrow$B for an average ion $\mathbf{\Lambda_{n_{e}}}$ are the first- and second-order moments respectively
\begin{eqnarray}\label{eq:positionwidth}
&&E^{\mathbf{\Lambda_{n_{e}}}}_{ab} = \mu_{1} =\frac{\sum_{\mathbf{K_{n_{e}}}}E^{\mathbf{K_{n_{e}}}}_{ab}\mathbf{w}_{ab}}{\mathbf{W}}  \nonumber  \\
&&(\Delta E^{\mathbf{\Lambda_{n_{e}}}}_{ab})^{2} = \mu_{2} =\frac{\sum_{\mathbf{K_{n_{e}}}}( E^{\mathbf{K_{n_{e}}}}_{ab}-E^{\mathbf{\Lambda_{n_{e}}}}_{ab})^{2}\mathbf{w}_{ab}}{\mathbf{W}}
\end{eqnarray}
Here $\sum_{a,b}$ is replaced by $\sum_{\mathbf{K_{n_{e}}}}$ since the array for an average ion includes all the configurations of the same number of bound electrons, i.e., $\mathbf{n_{e}}$.

The zero- and first-order configuration-average energies are\cite{STA1989}
\begin{eqnarray}\label{eq:EOFC}
&&E^{\mathbf{K_{n_{e}}}(0)}= \sum_{s}w_{s} \varepsilon_{s}    \nonumber  \\
&&E^{\mathbf{K_{n_{e}}}(1)}= \sum_{s}w_{s}\langle s \rangle + \frac{1}{2}\sum_{r,s}w_{s}(w_{r}-\delta_{rs})\langle r,s \rangle \nonumber
\end{eqnarray}
Here $w_{s}$ and $\varepsilon_{s}$ are the orbital occupation number and orbital energy. $\langle s \rangle$ and $\langle r,s \rangle$ are respectively the one- and two-electron matrix elements shown in Appendix B of Ref.\cite{STA1989}.

The zero- and first-order configuration-average transition energies for a$\rightarrow$b transition are
\begin{eqnarray}\label{eq:DEOFC}
&&E^{\mathbf{K_{n_{e}}}(0)}_{ab}= \varepsilon_{b} - \varepsilon_{a}   \nonumber \\
&&E^{\mathbf{K_{n_{e}}}(1)}_{ab}= D_{0} + \sum_{s}(w_{s}-\delta_{sa})D_{s} \nonumber
\end{eqnarray}
Here $D_{0}=\langle b \rangle - \langle a \rangle$ and $D_{s}=\langle s,b \rangle - \langle s,a \rangle$.

If we put $E^{\mathbf{K_{n_{e}}}(0)}_{ab}$ into Eq.(\ref{eq:positionwidth}), we get $E^{\mathbf{\Lambda_{n_{e}}}(0)}_{ab}= \varepsilon_{b} - \varepsilon_{a} $ and
$(\Delta E^{\mathbf{\Lambda_{n_{e}}}(0)}_{ab})^{2} = 0$.

In the present work, we use the following formulas to calculate the position and width of the array, which are derived from formulas (35) and (36) of Ref.\cite{STA1989} by using the first-order configuration energy $E^{\mathbf{K_{n_{e}}}(1)}_{ab}$.
\begin{widetext}
\begin{eqnarray}\label{eq:width-first}
&&E^{\mathbf{\Lambda_{n_{e}}}(1)}_{ab}= D_{0} + \sum_{s}\frac{\mathbf{\Omega}^{\mathbf{\Lambda_{n_{e}}}}_{s}}{g_{s}}g_{s}^{ab}D_{s} \nonumber \\
&&(\Delta E^{\mathbf{\Lambda_{n_{e}}}(1)}_{ab})^{2} = \sum_{t,r}(\frac{\mathbf{\Omega}^{\mathbf{\Lambda_{n_{e}}}}_{t}}{g_{t}}g_{t}^{ab}D_{t})(\frac{\mathbf{\Omega}^{\mathbf{\Lambda_{n_{e}}}}_{r}}{g_{r}}g_{r}^{abt}D_{r})+
\sum_{t}\frac{\mathbf{\Omega}^{\mathbf{\Lambda_{n_{e}}}}_{t}}{g_{t}}g_{t}^{ab}D^{2}_{t}-(E^{\mathbf{\Lambda_{n_{e}}}(1)}_{ab}-D_{0})^{2}
\end{eqnarray}
\end{widetext}
Here $g_{t}^{ab}=g_{t}-\delta_{at}-\delta_{bt}$ and $g_{r}^{abt}=g_{r}-\delta_{ar}-\delta_{br}-\delta_{tr}$.

Comparatively, in AA model, the summation of the orbital rate equations are carried out on all the configurations at all ionic stages and the position and width of the transition array are
\begin{eqnarray}\label{eq:widthAA}
&&E^{(AA)}_{ab} = D_{0} + \sum_{s}\mathbf{p}_{s}g_{s}^{ab}D_{s}                            \nonumber \\
&&(\Delta E^{(AA)}_{ab})^{2} = \sum_{t}\mathbf{p}_{t}(1-\mathbf{p}_{t})g_{t}^{ab}D^{2}_{t}
\end{eqnarray}
Here $\mathbf{p}_{s}$ is the occupation probability of orbital $s$, i.e., the ratio of the orbital occupation number to the statistic weight.

\section{Results and discussions}

In this work, 65 single orbital bases $nlj(n\leq5)$ and $nl(5<n\leq10)$ are used (listed in paper I).

\begin{figure}
\includegraphics[scale=0.35]{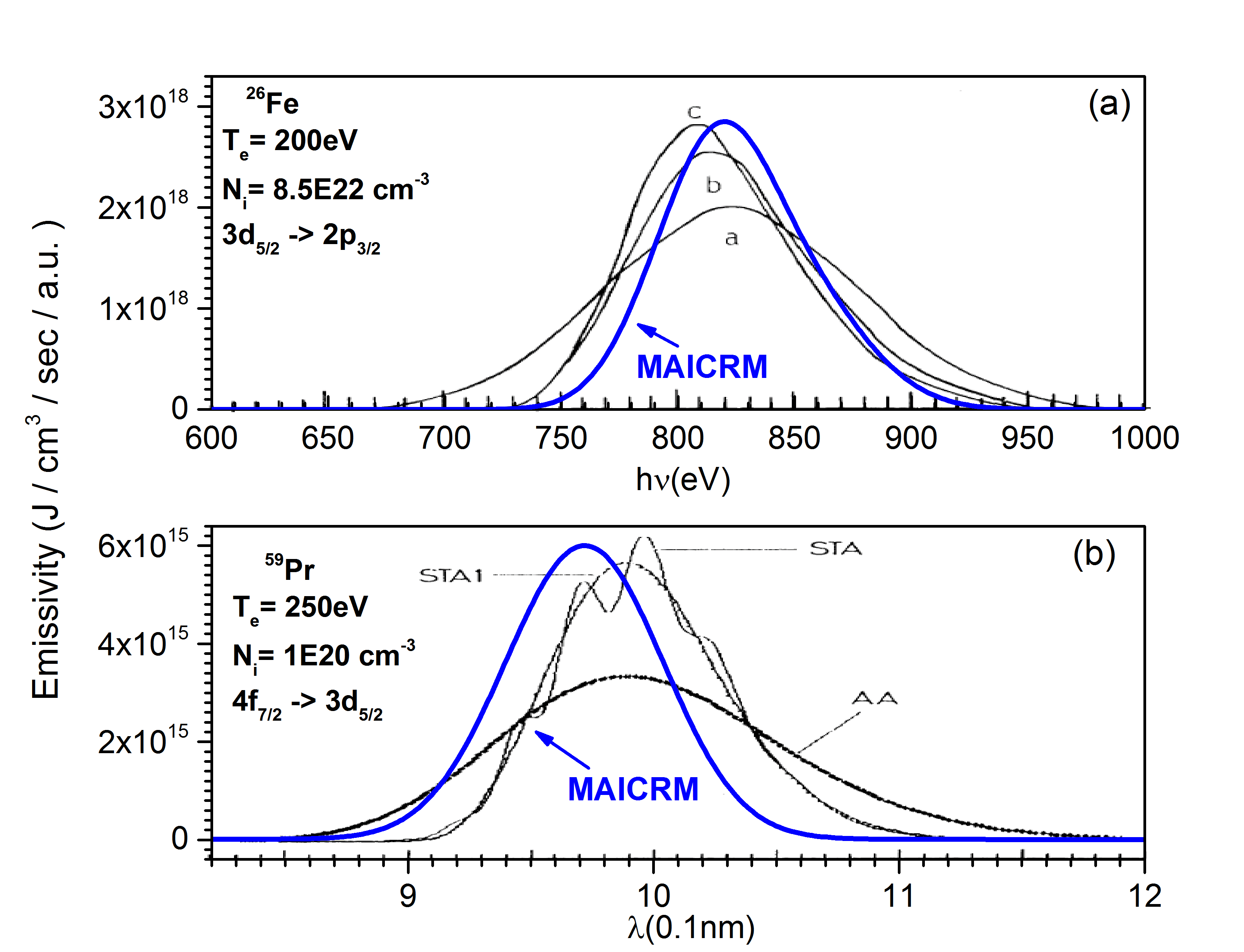}
\caption{\label{fig:FePr}(Color online) The comparisons of the one-electron transition emission spectra for Fe and Pr plasmas. The emissive intensity from Ref.\cite{STA1989} is in arbitrary units. (a) Emission spectra for the one-electron transition $3d_{5/2}-2p_{3/2}$ of Fe plasma at $T=$200eV and $N_{i}=8.5\times10^{22}$ cm$^{-3}$. Curve a : AA result (zeroth-order energies); curve b : STA result with a single supershell for each charge state and zeroth-order energies; and curve c : STA result with a single supershell for each charge state but using first-order energies. (b) Emission spectra for the one-electron transition $4f_{7/2}-3d_{5/2}$ of Pr plasma at $T=$250eV and $N_{i}=10^{20}$ cm$^{-3}$. Curve of AA : AA method; curve of STA : convergence with increasing number of STA; curve of STA1 : single supershell for each charge state. }
\end{figure}

Fig.\ref{fig:FePr} shows the comparisons of the one-electron transition emission spectra for Fe plasma at $T$=200eV, $N_{i}=8.5\times10^{22}$ cm$^{-3}$ and Pr plasma at $T$=250eV, $N_{i}=10^{20}$ cm$^{-3}$. The emissive intensity of Ref.\cite{STA1989} is in arbitrary units. As shown in Fig.\ref{fig:FePr}, compared with the STA curves, which are of a single supershell for each charge state, the widths of MAICRM curves are in good agreement, while the line positions show about $2\%$ red shifts, namely about 11eV for Fe and about 0.2${\mathbf{\AA}}$ for Pr. The shifts of the positions should result from the difference of the ionization distribution and the atomic data including the orbital energies and transition matrix elements. The widths of the AA curves are obviously broader than MAICRM and STA. From the statistic angle, the spectra of MAICRM and STA with a single supershell are both the function of the average orbital occupations of the configurations at one charge stage while the average orbital occupations are determined by different ways, namely by solving the excitation/de-excitation rate equations for MAICRM and by assuming a Boltzmann distribution for STA.

\begin{figure}
\includegraphics[scale=0.35]{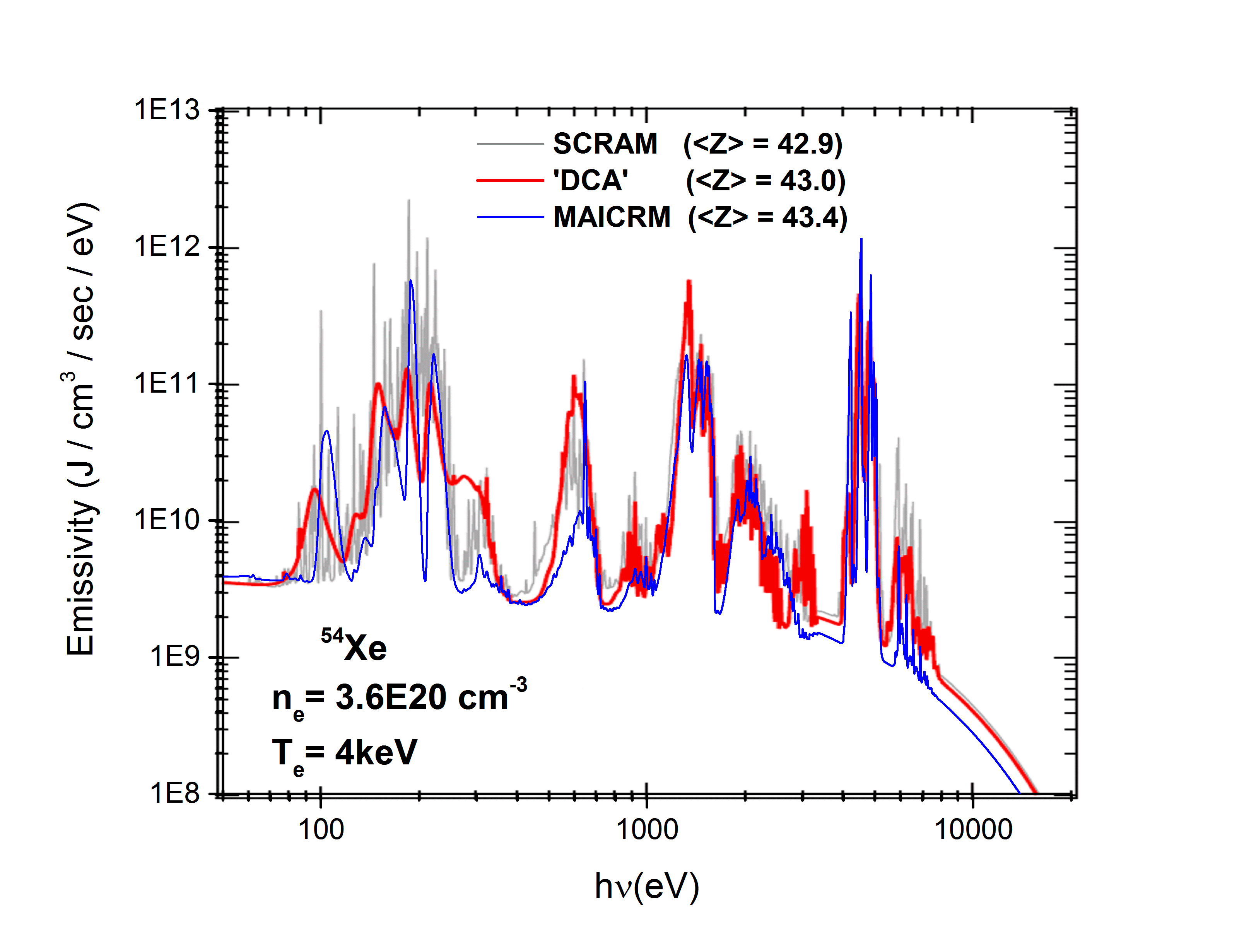}
\caption{\label{fig:Xe}(Color online) The calculated emissivity as a function of photon energy for Xe plasma at $T_{e}$=4keV and $N_{e}=3.6\times10^{20}$ cm$^{-3}$. The results of 'DCA' and SCRAM are from Ref.\cite{Scott2010}. }
\end{figure}

Fig.\ref{fig:Xe} shows the comparison of the emissivity of Xe plasma at $T_{e}$=4keV and $N_{e}=3.6\times10^{20}$ cm$^{-3}$. The detailed gray curve is calculated by SCRAM codes which uses hybrid atomic data, including levels, configurations and superconfigurations, constructed from a combination of fine structure and unresolved transition array (UTA) data from FAC, supplemented with the extended screened-hydrogenic model (including satellite shifts)\cite{SCRAM2007}. The red curve is calculated by extended 'DCA' codes which construct atomic levels as superconfigurations or Layzer complexes by shells described by principal quantum numbers with two extensions: one is to split each photoexcitation transition between principal quantum numbers into multiple term-to-term transitions with individual energies and oscillator strengths; the other one is to assign an additional width to each transition, representing additional unresolved fine structure multiplets, modeling each transition as an UTA\cite{Scott2010}. Here SCRAM and 'DCA' use a set of similar superconfigurations for each ionic stage, thus their calculated average ionizations $<Z>_{\mathbf{SCRAM}}=42.9$ and $<Z>_{'\mathbf{DCA}'}=43.0$ agree well. $<Z>_{\mathbf{MAICRM}}=43.4$ is higher by about 0.5.

In Fig.\ref{fig:Xe} the mean features of the three curves are in general agreement although MAICRM and extended 'DCA' have no the fine structures of SCRAM. Compared to the detailed gray spectrum of SCRAM, MAICRM curve agrees better than 'DCA' curve at some places such as the peaks at 310eV and the relative intensities of three peaks at 155eV, 190eV and 220eV. The peak of MAICRM curve at 105eV is closer to the maximum position of gray curve than the red curve. The continue background of MAICRM at energies larger than 1keV is lower than the grey and red curves which should result from the difference of $<Z>$.

\begin{figure}
\includegraphics[scale=0.35]{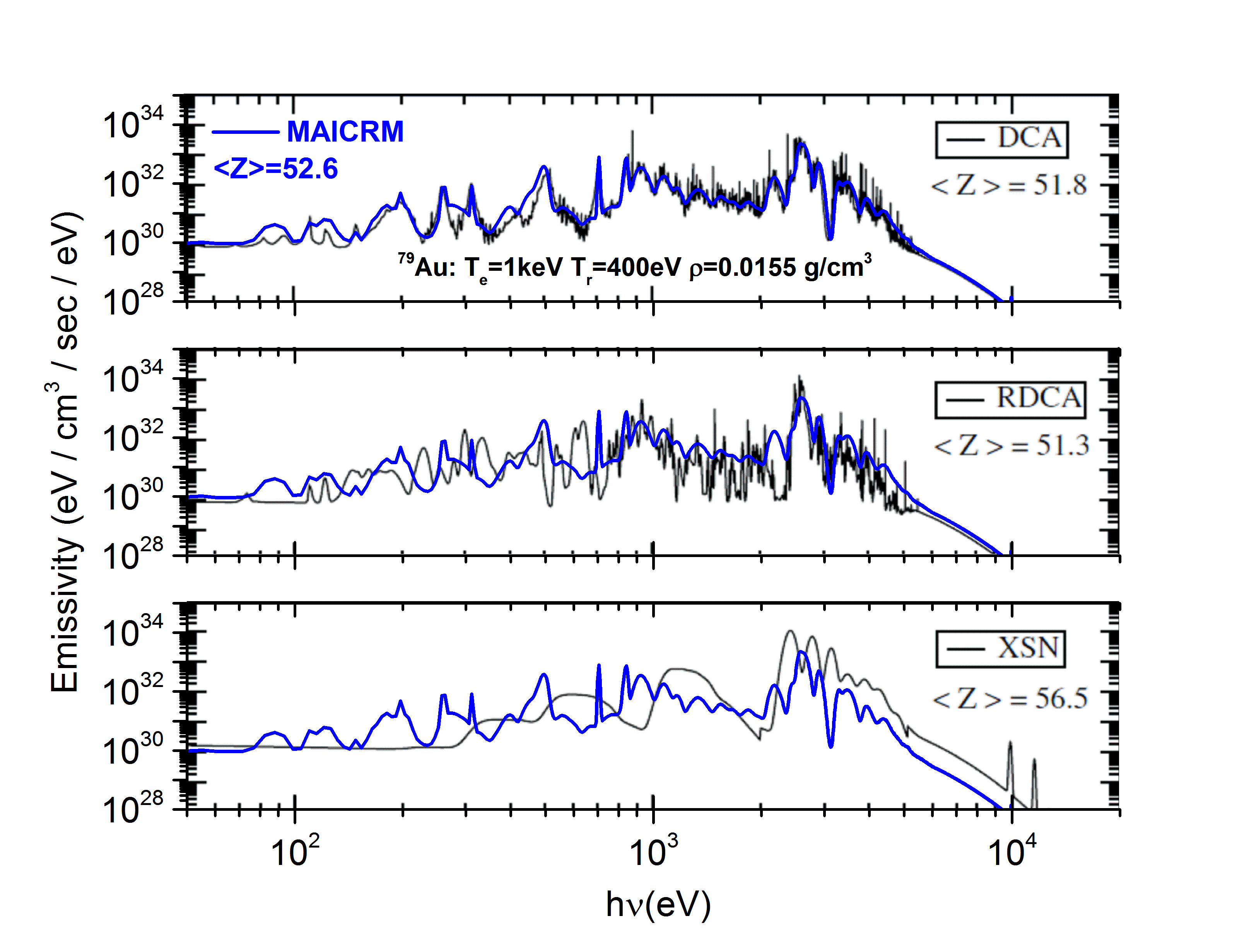}
\caption{\label{fig:Au}(Color online) The calculated emissivity as a function of photon energy for Au plasma at $T_{e}$=1keV, $T_{r}$=400eV and a mass density of $0.0155$ g/cm$^{3}$. For clear comparison with the results of DCA, RDCA and XSN of Ref.\cite{RDCA2009}, the blue curve of MAICRM is added into the three panels.}
\end{figure}

Fig.\ref{fig:Au} shows the comparison of the emissivity of Au plasma at $T_{e}$=1keV, $T_{r}$=400eV and $\rho$=$0.0155$g$/$cm$^{3}$. Here DCA is the traditional detailed configuration accounting model, RDCA is a reduced DCA model and XSN is a AA model\cite{Scott2010}. RDCA model makes an average of the energies and cross-sections of the detailed states falling within a spectral energy grid divided according to the DCA energy level structure of an ion\cite{Scott2010}. Compared with $<Z>_{\mathbf{DCA}}$=51.8, $<Z>_{\mathbf{MAICRM}}$=52.6 is much closer than $<Z>_{\mathbf{XSN}}$=56.5. For clarity, the blue curve of MAICRM is added into the three panels. Panel (a) shows the MAICRM curve agrees well with the DCA curve especially at energies larger than 1keV. In the energy range of 80eV-120eV, the widths of MAICRM curve are larger than DCA. Panel (b) shows the RDCA curve differs from the MAICRM curve in many places. In panel (c) the XSN curve shows simpler structure and broader peaks than MAICRM since in AA model only one average ion is considered.

\section{Conclusion}
We have developed a general model to simulate the plasma properties of hot dense plasma. The average orbital occupations and total population for the configurations within one charge state are characterized by an average ion. The orbital occupation $\{ \mathbf{\Omega}^{\mathbf{\Lambda_{n_{e}}}}_{n}\}$ and population $\{P_{\mathbf{\Lambda_{n_{e}}}}\}$ of the average ion $\mathbf{\Lambda_{n_{e}}}$ are obtained by solving two sets of rate equations sequentially and iteratively. Our calculated CSDs as well as the spectra of hot dense plasmas agree with DCA/SCA models as shown in paper I and this work. On the other hand, since MAICRM only considers one configuration at each charge state the computational cost reduces magnitudes compared to DCA/SCA models due to their hundreds or thousands atomic levels at one charge state, which means the possibility to be coupled into radiation-transport hydrodynamic simulations in the future.

\begin{acknowledgments}
This work is partly supported by the National Key R\&D Program of China Under Grant No. 2017YFA0402300.
\end{acknowledgments}

\nocite{*}

\end{document}